# Proton transport through nanoscale corrugations in two-dimensional crystals

O. J. Wahab[1], E. Daviddi[1], B. Xin[2, 3], P. Z. Sun[2, 3], E. Griffin[2, 3], A. W. Colburn[1], D. Barry[2], M. Yagmurcukardes[4], Francois Peeters[5], A. K. Geim[2, 3], M. Lozada-Hidalgo[2, 3], P. R. Unwin[1]

[1] Department of Chemistry, University of Warwick, Coventry CV4 7AL, United Kingdom
[2] Department of Physics and Astronomy, The University of Manchester, Manchester M13 9PL, UK
[3] National Graphene Institute, The University of Manchester, Manchester M13 9PL, UK
[4] Department of Photonics, Izmir Institute of Technology, 35430, Izmir, Urla, Turkey
[5] Departement Fysica, Universiteit Antwerpen, Groenenborgerlaan 171, B-2020 Antwerp, Belgium

**Defect-free graphene is impermeable to all atoms[1–5] and ions[6,7] at ambient conditions. Experiments that can resolve gas flows of a few atoms per hour through micrometre-sized membranes found that monocrystalline graphene is completely impermeable to helium, the smallest of atoms[2,5]. Such membranes were also shown to be impermeable to all ions, including the smallest one, lithium[6,7]. On the other hand, graphene was reported to be highly permeable to protons, nuclei of hydrogen atoms[8,9]. There is no consensus, however, either on the mechanism behind the unexpectedly high proton permeability[10–14] or even on whether it requires defects in graphene's crystal lattice[6,8,15–17]. Here using high resolution scanning electrochemical cell microscopy (SECCM), we show that, although proton permeation through mechanically-exfoliated monolayers of graphene and hexagonal boron nitride cannot be attributed to any structural defects, nanoscale non-flatness of 2D membranes greatly facilitates proton transport. The spatial distribution of proton currents visualized by SECCM reveals marked inhomogeneities that are strongly correlated with nanoscale wrinkles and other features where strain is accumulated. Our results highlight nanoscale morphology as an important parameter enabling proton transport through 2D crystals, mostly considered and modelled as flat, and suggest that strain and curvature can be used as additional degrees of freedom to control the proton permeability of 2D materials.**

Measurements of proton transport through 2D crystals demonstrated that these pose an energy barrier for incoming protons of ~0.8 eV and ~0.3 eV for graphene and hexagonal boron nitride (hBN), respectively[8]. Additional experiments with hydrogen's heavier isotope deuterium revealed that the initial energy of incoming protons is not given by thermal excitations (~25 meV) but raised by ~0.2 eV due to zero-point oscillations of protons bound to oxygen atoms in the proton conductive media[9]. This correction lifts the total energy barriers $E$ posed by the crystals to ~1.0 eV and ~0.5 eV for graphene and hBN, respectively. Despite these insights, the mechanism for proton permeation through the 2D crystals remains controversial. The general consensus from density functional theory (DFT) calculations is that the energy barriers should be notably larger[14]. The studies (e.g., refs. [10,11,13,14,18]) have yielded a rather wide range of $E$ but always exceeding ~1 eV found experimentally. The spread of values arises from the various assumptions made in the models, such as whether the process is slower than the lattice relaxation timescale[14], protons tunnel quantum-mechanically[11,12] or protons locally hydrogenate the carbon lattice (and hence locally expand it) prior to transfer[13,19]. This uncertainty has motivated an alternative explanation widely speculated in the literature, namely that proton permeation takes place through structural defects in the crystal lattice. This hypothesis is based on experiments using graphene grown by chemical vapour deposition (CVD)[15–17], which has grain boundaries, pinholes and other imperfections that appear during growth and transfer[20–22].

Experiments using CVD graphene typically report very high proton permeation rates and, sometimes, even the loss of graphene's impermeability to other ions[16]. However, the explanation that assumes atomic-scale defects as the only proton conductive sites is inapplicable to mechanically-exfoliated graphene. Indeed, transmission and tunnelling electron microscopy have failed to observe any vacancies or other atomic-scale imperfections for scans over relatively large areas of such crystals. Even more decisively, gas permeation experiments that can easily detect a single angstrom-scale defect permeable to gases within micrometre-sized membranes[1,2,4,5], detected none in exfoliated graphene and hBN monolayers[6]. Further experimental evidence is necessary to understand proton transport through defect-free 2D crystals and resolve the existing controversy.

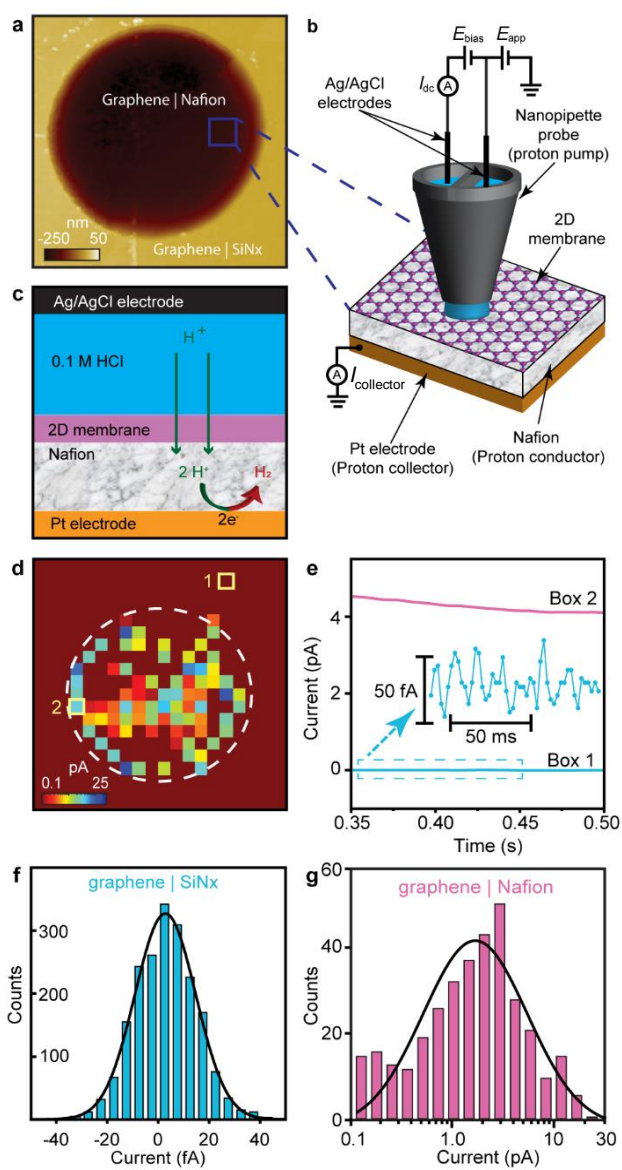

**Figure 1 | Nanoscale visualization of proton currents through 2D crystals. a,** AFM image of one of our experimental devices, showing the circular aperture in the SiN$_x$ substrate covered with monolayer graphene. The bottom side of the membrane is in contact with Nafion (proton conducting polymer) and the top side is left accessible to SECCM. **b**, Schematic of our SECCM setup. Blue box in **a** indicates the area zoomed-out into panel **b**. The nanopipette has two reservoirs filled with HCl electrolyte, each

electrically connected using Ag/AgCl electrodes. Potential $E_{bias}$ drives a current between these two electrodes ($I_{dc}$). The current acts as a feedback signal that detects if the probe is in contact with a sample surface. After a contact is established, the potential $E_{app}$ is used to pump protons from the HCl reservoirs through graphene and onto the Pt electrode. This yields $I_{collector}$, which is the current reported in the SECCM maps. The size of each measured pixel is given by the size of the nanopipette tip and the formed meniscus. Pixelated measurements are repeated over extended areas to generate a map ('SECCM scanning protocol' in Methods). **c**, Schematic of proton flow through the devices. Protons are injected from the nanopipette through graphene into Nafion. After reaching the Pt proton collector, they evolve into $H_2$ gas. **d**, Example of SECCM maps ($I_{collector}$ maps) for apertures covered with graphene. The white dotted circle marks the 2 μm diameter aperture in $SiN_x$. Colour scale bar, current in pA. **e**, Steady-state SECCM currents as a function of time for graphene-on-Nafion (pink) and graphene-on-$SiN_x$ (blue). The measurements are from the areas indicated by boxes in **d** (colour coded). **f**, Statistical distribution of the measured values of steady-state proton currents for graphene-on-$SiN_x$. The peak-to-peak noise level is ~50 fA (inset of **e**), which reduces to ~10 fA after time averaging. **g**, Similar statistics for graphene-on-Nafion. The SECCM currents were measured over the entire area shown in the maps, including the regions shown as brown pixels for both graphene-on-Nafion and graphene-on-$SiN_x$ (see panel **d**). In both panel **f** and **g**, each count represents the average of steady-state current over 100 ms. Data are collected from 6 different devices. Solid curves, best Gaussian fits.

In this report, we investigate the distribution of proton currents through mechanically-exfoliated 2D crystals with high spatial (nanoscale) and high current (fA) resolution using scanning electrochemical cell microscopy (SECCM). The devices for this study consisted of graphene and hBN monolayer crystals, which were suspended over micrometre-sized holes (2 μm in diameter) etched into silicon-nitride ($SiN_x$) substrates (Extended Data Fig. 1; Methods). No structural defects are expected in the 2D crystals, as dozens of similar membranes were studied in ultra-sensitive gas flow experiments, with none showing any permeation of helium[2] (see 'Absence of defects in mechanically exfoliated 2D membranes' in Methods). One side of the obtained free-standing 2D membranes was coated with a proton conducting polymer (Nafion) that was in turn electrically connected to a mm-sized Pt electrode. The opposite side of the 2D crystal was left exposed to air and probed using SECCM (Extended Data Fig. 2). For SECCM measurements, a nanopipette with a tip opening diameter of ~200 nm and filled with 0.1 M HCl was accurately positioned over the sample using piezo drivers (Figs. 1a&b). Upon contact with the surface, a droplet meniscus was formed, whose size determines the surface area being probed (Extended Data Figs. 2&3). During such measurements, protons from the HCl reservoir in the pipette are injected through the sample with the potentials $E_{app}$ and $E_{bias}$ set so as to fix $E_{collector}$ = -0.5 V at the Pt electrode ($H^+$ collector) with respect to Ag/AgCl ('SECCM scanning protocol' in Methods). Therefore, the 2D crystal constitutes an atomically thin barrier between the SECCM probe ($H^+$ pump) and the Nafion-Pt collector, and a current ($I_{collector}$) is detected only when the probe is at locations where $H^+$ transmission occurs (Fig. 1). This barrier is the current-limiting element in our devices, as corroborated directly by measuring the SECCM response at areas without 2D crystals (bare Nafion), which give currents >3 orders of magnitude higher (Extended Data Fig. 2).

In the SECCM measurements, we acquire current vs time curves for each spatial location tested. These curves display resistor-capacitor decay characteristics and a steady state is achieved typically within ~400 ms after meniscus formation at the sample surface (Fig. 1 and Extended Data Fig. 2). All the SECCM maps presented below are in the steady state. Figure 1d shows an example of such maps obtained from monolayer graphene. If the device is scanned over areas of graphene covering the $SiN_x$ substrate, only small parasitic (leakage) currents of ~10 fA are observed because the $SiN_x$ substrate blocks proton transport (Figs. 1e&f). In contrast, for the areas where graphene is in direct contact with Nafion, proton currents of up to several pA are observed. Intriguingly, the SECCM maps (Fig. 1d, Extended Data Fig. 4) show that proton transport through graphene is spatially highly inhomogeneous, and this was the case for all the studied devices (more than 20). While several pixels inside graphene-on-Nafion areas show currents within our background noise, statistics for the other pixels display a log-normal distribution with its mode located at ~2 pA, two orders of magnitude above the noise level (Fig. 1g).

It is instructive to compare these results with measurements on similar devices but made from CVD graphene. Previously[15–17], proton transport through CVD graphene was attributed to sparsely distributed defects (probably microholes; one per $10^3$–$10^4$ μm$^2$), each displaying a proton current[16,17] of ~0.3–1 nA under similar $E_{bias}$. Our present SECCM measurements using higher-quality CVD graphene have not found such isolated highly-conductive defects (possibly pinholes) and instead show enhanced permeation over large areas, mostly at what seems to be grain boundaries, a common feature of CVD-grown graphene[7,20] (Extended Data Fig. 5, 'SECCM characterisation of CVD graphene membranes' in Methods). The SECCM currents display a log-normal distribution with the mode (peak) at ~20 pA. This is 2 orders of magnitude lower than the current through individual defects found previously in CVD graphene but an order of magnitude higher than that in our mechanically exfoliated graphene monocrystals. In this context, exfoliated graphene is fundamentally different from CVD graphene. Indeed, neither nanoscale holes nor grain boundaries are present in our monocrystals but the experimental resolution still allows us to observe ~100 proton-conductive sites per μm$^2$. Because of the proven gas and ion impermeability of mechanically exfoliated graphene[2,5], these sites cannot be attributed to atomic-scale defects, not even vacancies[5]. Accordingly, we must conclude that a defect-free graphene lattice is proton permeable, in agreement with earlier conclusions[8,9]. In the rest of this report, we explore the origins of the unexpected spatial inhomogeneity of proton transport through defect-free monolayers of graphene and hBN.

To understand the observed spatial inhomogeneity of SECCM maps, we compare them with atomic force microscopy (AFM) images of the 2D crystals. Figs. 2 a-d show AFM adhesion force maps and corresponding SECCM scans for two graphene devices (for more examples, see Extended Data Fig. 4). The AFM micrographs reveal that the membranes are not flat but contain wrinkles that are a couple of nm in height $h$ and tens of nm in width $L$ ($h/L \approx 0.06$-$0.18$, as found from topography maps; Extended Data Fig. 4). It is clear from Figs. 2a-d that the positions of the wrinkles closely correlate with some of the most highly conductive regions in the SECCM maps (blue pixels). Other areas of high proton conductivity occur around the apertures' rims. The common denominator for the two types of high-conductivity regions is that, in both cases, the 2D membrane is under notable strain. While the whole membrane is strained for being suspended, the stress mainly accumulates around the rim[23]. The resultant tensile strain is estimated at a few percent ('Atomic force microscopy characterisation' in Methods). Stress is also known to accumulate near wrinkles[24], which can result in strain similar to that around the aperture's rim.

Next, we describe similar experiments with devices fabricated using monolayer hBN instead of graphene. Fig. 2e shows one of our hBN devices, in which half a SiN$_x$ aperture is covered with monolayer hBN and the other half with tetra-layer (additional examples in Extended Data Fig. 6). The areas covered with the tetra-layer are notably flatter than those with monolayer and display no proton transport within our resolution, even at high-strain areas around the rim. This is in agreement with our previous work where no proton permeability could be detected for ≥4 layers of hBN[8]. In stark contrast, areas covered with monolayer hBN display a high density of highly conductive sites with currents generally larger than in graphene devices. This is consistent with the fact that hBN monolayers are on average ~50 times more proton conductive than graphene[8,9]. As in graphene devices, the highest activity (blue pixels) in the SECCM map is concentrated around wrinkles and the rim. However, the maps for hBN monolayers also reveal many active areas that do not correspond to any obvious morphological features. In the statistical distribution (Fig. 2h), the corresponding currents result in a notable shoulder centred at ~10 pA, whereas currents from wrinkled areas are centred at ~50 pA. This allows us to estimate that wrinkles accelerate proton transport by a factor of ~5 with respect to that from featureless regions. Although no such shoulder was apparent for graphene membranes, some SECCM regions without morphological features also exhibited many active pixels with currents ~0.1-1 pA (well above the noise level). This may suggest a similar relation between current amplitudes for strained and featureless regions in graphene membranes.

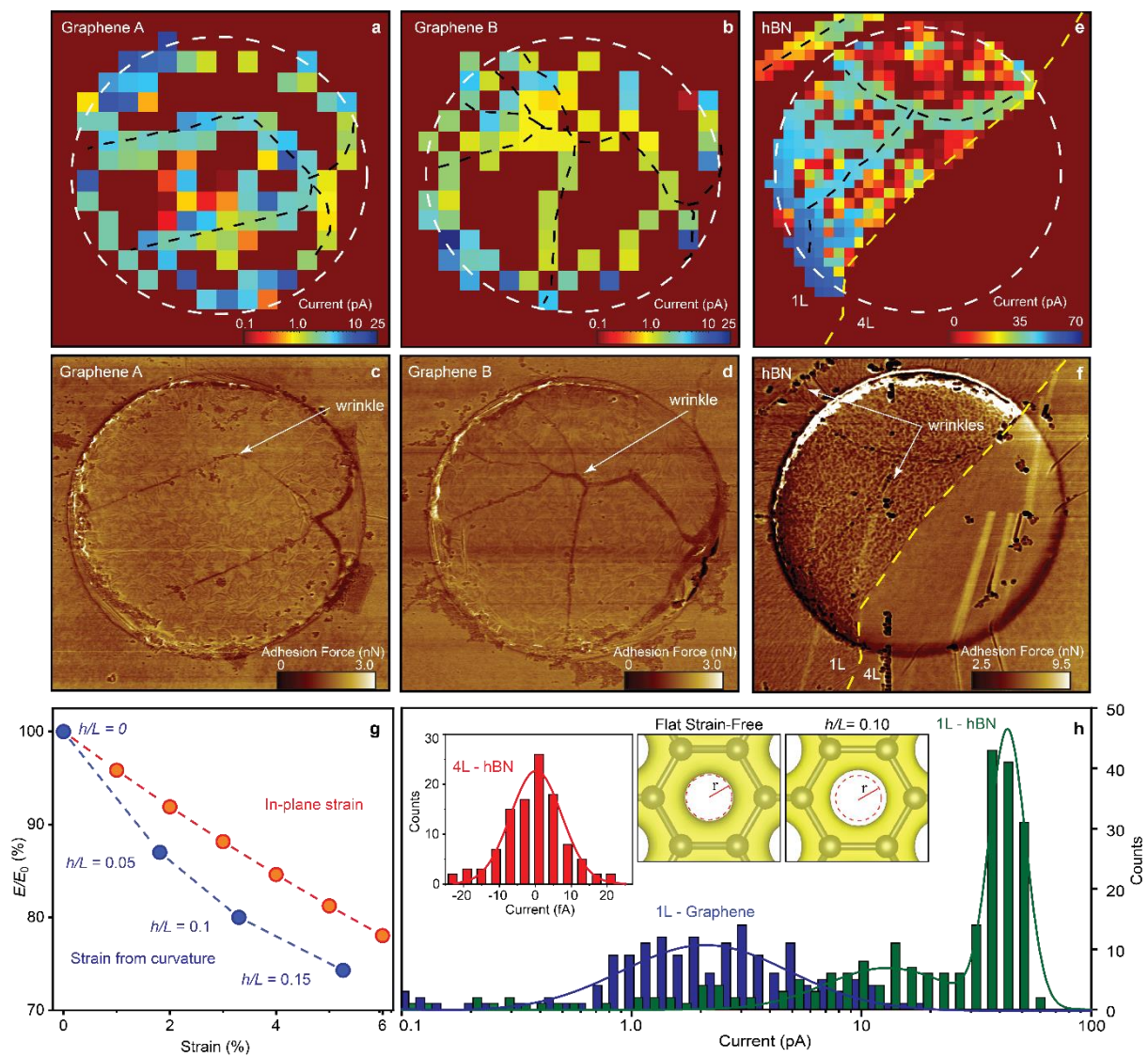

**Figure 2 | Unexpected inhomogeneity of proton transport through 2D crystals**. **a**, **b**, SECCM maps for two graphene devices. The white circles mark the rim of the apertures in SiN$_x$. **c**, **d**, AFM force maps for the devices in the panels above. Wrinkles and edges are clearly visible in the AFM maps and correlate with high-conductivity areas in the SECCM maps. For easier comparison, the black dashed curves in **a** and **b** mark wrinkles' positions. **e**, Proton currents through an hBN device. Yellow dashed

curve, border between monolayer (1L; left) and tetra-layer (4L) hBN (right). **f**, AFM force map for the above device. Apparent wrinkles are indicated by the arrows and marked by the black curves in **e**. An interesting feature of this particular device is notable proton currents in the top left corner in **e**, away from the aperture in $SiN_x$. Extended Data Fig. 6 reveals that this feature is due to a wrinkle originating from a neighbouring aperture. The wrinkle provides a nanocavity between hBN and the $SiN_x$ substrate, which allows protons to reach this area. **g**, Strain lowers the energy barrier $E$ for proton permeation ($E_0$ is the barrier for unstrained graphene). Blue symbols, the effect of strain arising from curvature; values of $h/L$ are specified next to each point. Red data, $E/E_0$ due to purely in-plane strain. **h**, Statistics of proton currents for graphene and hBN monolayers (data from panels **a**, **b**, **e**). Left inset, statistics collected from the tetra-layer region. Solid curves, best Gaussian and double-Gaussian fits for graphene and 1L-hBN, respectively (accuracy of ~10% in determining the modes of the normal distributions). The right inset shows the calculated electron density provided by the crystal lattice for unstrained (left) and strained (right) graphene; the latter calculations are for strain arising from curvature with $h/L$ = 0.10. To make changes in the electron density evident, the dashed red circle in the left panel marks the boundary between regions[8] with densities above and below 0.2 $e$ Å$^{-3}$ (the latter region is shown in white). The same circle is projected onto the right panel and emphasizes that the low-density region expanded in the strained lattice.

We attribute the smaller proton currents away from apparent morphological features to transport through nanoscale ripples that are ubiquitous in 2D crystals[2,25–27]. These ripples can be either dynamic (flexural phonons in freely suspended membranes) or static (caused by strain or adsorbates), as previously revealed by transmission electron microscopy[26]. While our AFM maps do show a difference in apparent roughness for mono- and tetra- layer regions in Fig. 2f that can be attributed to static ripples, these ripples are too small ($L$ of several nm[25,26]) to be quantified using our AFM. Nevertheless, direct evidence from transmission electron microscopy has previously found[25,26] that static ripples have typical $h/L \approx 0.1$, comparable to the aspect ratios observed for the wrinkles. On this basis, we propose that unavoidable nanorippling of 2D crystals enhances their proton permeability in much the same way as larger wrinkles. Because nanoripples have smaller $L$, their smaller areas result in smaller proton currents within individual pixels. To corroborate this microscopic picture provided by SECCM, we integrated the observed currents over the entire area of the 2D membranes. This allows for an estimate of the proton conductivities of graphene and hBN monolayers as ~10 mS cm$^{-2}$ and ~300 mS cm$^{-2}$, respectively. Taking into account our experimental uncertainty (factor of ~3 that comes mostly from assessing the area contacted by the probe during the scans), the found conductivities are in good agreement with the previous measurements of average proton conductivities of the 2D crystals[8].

Our DFT calculations (Methods) provide further support for the above explanation of localised strain as the main reason for the observed spatial inhomogeneity in proton transport. Indeed, the barriers imposed by the 2D crystals for incoming protons depend on the density of electron clouds associated with the crystal lattice[8]. For example, hBN monolayer presents a sparser electron density than graphene and, accordingly, is more permeable to protons[8]. Strain and curvature modify the electron-density distribution within 2D materials (Fig. 2h, inset), and this can enhance their transparency to protons. Our calculations show that, by stretching the graphene lattice by ~5%, the energy barrier $E$ for protons is lowered by ~20% (Fig. 2g, Extended Data Fig. 7). If this strain is accompanied by curvature (like in the case of ripples), the barrier is reduced further, so that $E$ reaches ~75% of the unstrained value $E_0$ (Fig. 2g). Although this reduction seems relatively small, proton currents depend exponentially on the barrier height[8], which means that proton transport can be accelerated by several orders of magnitude within strained regions around wrinkles, ripples and other morphological features.

In conclusion, our experiments show that strain-inducing morphological features in otherwise defect-free 2D crystals are associated with enhanced proton conductivity around them. A notable example of this is graphene wrinkles that do not require any crystal-lattice defects but result in high proton currents, not dissimilar to the case of grain boundaries in CVD graphene (Methods). Our findings also suggest that nanoscale ripples, ubiquitous in 2D membranes and known to result in considerable strain, accelerate proton transport within nominally flat areas. This is important as graphene is typically modelled as a perfectly flat unstrained crystal. Because strain and curvature in 2D membranes can typically reach up to 10%, the theories predicting $E_0$ of up to 1.5 eV for flat unstrained graphene (e.g., refs[10,14]) appear to be consistent with the experiments reporting barriers of ~1.0 eV[8]. Finally, strain and curvature can be exploited to enhance proton conductivity of 2D crystals, which is of interest for various applications involving proton transport[9,28–30].

## Methods

**Device fabrication.** Micrometre-sized apertures were etched into silicon/silicon-nitride substrates (500 nm $SiN_x$) using photolithography, wet etching and reactive ion etching, following the protocol previously reported[8]. Our devices had several apertures next to each other and were 2 µm in diameter each (Extended Data Fig. 1). Monolayers of graphene and hBN were obtained by micromechanical cleavage[31] and identified using a combination of optical and atomic force microscopies and Raman spectroscopy, as previously reported[8,32,33]. The monolayers were suspended over the apertures in the $SiN_x$ substrate. The resulting free-standing membranes were coated on one side by drop-casting the Nafion polymer (5%, 1,100 equiv wt) to obtain ~10 µm thick films. The devices were annealed in a water-saturated environment at 130 °C to crosslink the polymer. The collector electrode was prepared by laminating a Pt foil (10 × 10 mm, 99.95% purity, Goodfellow Ltd, UK) onto a cylindrical carbon block with a hot compression mounting machine (SimpliMet, Buehler, USA). The exposed Pt surface was then subjected to mechanical and electrochemical polishing. Gaskets were used to cover the carbon block and expose only the Pt surface for contact with Nafion|2D crystal devices[34]. For measurements, the Nafion film was hydrated with deionized water and allowed to equilibrate before placing it in contact with the Pt collector.

**SECCM Probes.** Nanopipettes for SECCM were fabricated from quartz theta capillaries with filaments (WAR-QTF120-90-100, Friedrich & Dimmock, USA). The capillary (outer diameter, 1.2 mm; inner diameter, 0.90 mm; length, 100 mm) was pulled to form a fine sharp point with a tip opening diameter of ~200 nm, using a $CO_2$-laser puller (Sutter Instruments P-2000). The nanopipette was then filled with 100 mM HCl electrolyte, and a silicone oil layer was added on top of the electrolyte solution in the tip to minimise evaporation during prolonged scanning procedures[35]. Two AgCl-coated Ag wires, fabricated by electrochemically oxidising Ag wires (0.125 mm in diameter) in saturated KCl solution[36] were used as quasi-reference counter electrodes (QRCEs). Each of the nanopipette channels was fitted with a QRCE positioned about 3 - 4 cm away from the tip end[36].

**SECCM Instrumentation.** SECCM was performed using a home-built workstation[37]. The SECCM probe was mounted on a Z piezoelectric positioner (P-753.1CD LISA, Physik Instrumente, Germany) while the studied graphene or hBN device was mounted on a XY-piezoelectric positioner (P-622.2CD PIHera, Physik Instrumente). The SECCM probe was moved to the initial scan position using an XY-micropositioner (M-461- XYZ-M, Newport, USA) controlled with Picomotor Actuators (8303, Newport). An optical camera provided a visual guide for the probe positioning. The microscopy stage and all positioners were enclosed in a Faraday cage with heat sinks and vacuum panels to minimise noise and thermal drift. The Faraday cage was placed on an optical tabletop with tuned damping (RS 2000, Newport) balanced on a high-performance laminar flow isolator (S-2000 Series, Newport).
Data acquisition and instrumental control were carried out using an FPGA card (PCIe-7852R) running the Warwick Electrochemical Scanning Probe Microscopy software (www.warwick.ac.uk/electrochemistry/wec-spm). Two home-built electrometers were used for current measurements, together with home-built 8$^{th}$-order brick-wall filters with the time constant of the current amplifier set to 10 ms. Influence of acquisition parameters on the noise level is detailed elsewhere[38]. The current data were acquired every 4 µs and 256 samples were averaged to give a data acquisition rate of ~1 ms.

**Scanning protocol.** SECCM was deployed in the hopping mode[34,39,40] by recording spatially resolved current vs time (*I-t*) traces of the reduction of protons at the bottom Pt collector electrode (Extended Data Figure 2). The hopping mode protocol involved the approach of the probe to the 2D crystal surface until the electrolyte meniscus at the end of the tip (not the nanopipette itself) made contact (as detailed below). After a measurement on one site was completed, the probe was retracted and moved to the next site to generate a map of *I-t* traces over the entire device surface.

Two voltage controls are important in this protocol. The first is the potential difference, $E_{bias}$, between the quasi-reference counter electrodes in each of the two channels of the nanopipette (Extended Data Fig. 2a). This gives rise to an ion current between the two channels in the nanopipette, $I_{dc}$, which is used as a feedback signal to detect if the meniscus is in contact with the surface. When the droplet meniscus touched the device surface, a spike in $I_{dc} \gg 100$ pA signified 'jump-to-contact'[41] (detected with a feedback threshold of 45 pA; Extended Data Figs. 2b&c). When this threshold was reached, the probe motion was stopped. Note that this signal provided a means of landing the meniscus on the 2D crystal surface, irrespective of its local proton permeability. Further details of the $I_{dc}$ transients are provided in section 'Consistency of meniscus-surface wetting'.

The second voltage, $E_{app}$, between the nanopipette probe and the proton-collecting working electrode sets the potential of the Pt collector electrode with respect to the QRCEs as $E_{collector} = -(E_{app} + E_{bias}/2)$[42]. This potential is chosen as $E_{collector} = -0.5$ V vs. Ag/AgCl QRCE (equivalent to an overpotential of ~0.2 V vs the standard potential for the hydrogen evolution reaction). This was the maximum voltage used in ref.[8] where the linear-response proton transport was studied in similar devices. $E_{collector}$ drives the electrochemical proton reduction at the Pt electrode, which results in current $I_{collector}$ (inset of Extended Data Fig. 2). $I_{collector}$-*t* measurements were made for 500 ms and involved the area defined by the meniscus between the SECCM nanopipette and the substrate. After each measurement within this temporary and spatially localised droplet cell, the probe was retracted at a speed of 4 µm/s (Extended Data Fig. 2b) and moved to the next location where the above procedure was repeated. This allowed us to obtain a spatial- and time- resolved dataset for $I_{collector}$. Proton-current maps in our figures are presented as an average of the last 100 ms of the $I_{collector}$-*t* transients .

The z-position of the probe was recorded synchronously throughout the whole measurement procedure, with the value at the end of each nanopipette approach yielding a topographical map of the studied 2D-crystal device. Nonlinear sample-tilt and piezo-drift effects in such topographical maps were corrected using the scanning-probe-image-processing software package (v6.0.14, Image Metrology, Denmark). SECCM topographical maps of mechanically exfoliated graphene devices (not shown) were similar to those subsequently obtained by AFM, as expected. Such maps, recorded synchronously with proton transport activity, were especially valuable as they revealed morphology around proton-conducting sites in CVD graphene (Extended Data Figure 5).

**Consistency of meniscus-surface wetting.** The SECCM maps of proton transport through graphene and hBN monolayers display a marked spatial inhomogeneity. To establish that this is an intrinsic property of the 2D crystals and not a result of variations in surface-probe contact, we investigated the consistency of meniscus-surface wetting by analysing the current $I_{dc}$ flowing between the two channels in the nanopipette. Below we explain how $I_{dc}$ is used as a feedback signal that unequivocally detects meniscus-surface wetting, regardless of proton transport through a 2D crystal.

Extended Data Fig. 3 illustrates the steps that take place during the SECCM scan and how $I_{dc}$ changes during each step. Initially, the probe is not in contact with the sample (step *i*, 'approach') and $I_{dc}^i$ is constant as a function of time (~400 pA in this case). As the probe gets closer, the meniscus encounters the sample (step *ii*, 'meniscus touch'). In this step, $I_{dc}^{ii}$ sometimes decreased very slightly with respect to $I_{dc}^{i}$ ($\Delta I_{dc} = I_{dc}^{ii} - I_{dc}^{i} < 0$), attributed to slight squeezing of the meniscus. However, this depended on a specific 2D material measured. For graphene samples, we typically see a decrease of about 1% or 5 pA (Extended Data Figs. 3b&c), whereas for hBN we do not see such a drop (Extended Data Figs. 3e&f). We attribute this difference in behaviour to a stronger attraction of the electrolyte in the probe to hBN than graphene. The next step is meniscus wetting (step *iii*, 'meniscus wets'). This step is characterised by a sharp increase in current, $\Delta I_{dc} = I_{dc}^{iii} - I_{dc}^{i} > 200$ pA, which is an unmistakable indicator that the meniscus has fully wetted the sample (Extended Data Figs. 3b&d and 3e&g for graphene and hBN, respectively). The DC current then drops to a steady state (step *iv*) during which the meniscus stabilises. After the pre-programmed measurement period (500 ms of meniscus contact), the tip is retracted (step *v*, 'meniscus stretch' and step *vi*, 'meniscus detached'), with $I_{dc}$ first sharply increasing and then returning to the initial value. These steps were clearly visible throughout scanning of entire samples.

The described behaviour was observed independently of $I_{collector}$, that is, whether the proton current is being pumped or not through the device into the proton collector. Extended Data Figs. 3b&c also show that $I_{dc}$ displays the same features both in areas of high proton conductivity (blue curve) and in areas where no proton transport takes place (red). This shows that meniscus wetting of the sample is independent of proton transport through 2D crystals. Note, however, that the magnitude of $I_{dc}$ does change for active and inactive areas because the Ag/AgCl electrodes are also the counter electrodes for proton conductivity measurements[42]. This change served as independent confirmation of those sites where there was notable proton permeation through 2D crystals.

Note that the above also rules out changes in the droplet cell size as the source of the observed spatial inhomogeneity of the proton currents. The consistency of the SECCM cell size across the surface is also in accord with the following considerations. First, wrinkles protrude at most a few nanometres from flat areas of graphene, which leads only to small variations in the involved surface area as compared to the area probed in each pixel (ca. 200 nm in diameter). Therefore, this cannot explain the orders of magnitude difference in the observed SECCM activity. Second, the roughness associated with the wrinkles is much less than a typical surface roughness of a wide range of samples previously studied by SECCM for which consistent meniscus cell size was observed or deduced.[34,35,37,40,42–44]

**AFM and SEM characterisation.** High-resolution topography and adhesion AFM imaging was performed under ambient conditions using Bruker Dimension Icon AFM (Bruker, USA) employing the Peak Force mode. The instrument was equipped with SCANASYST-AIR silicon tips (Bruker, USA). The tips had a nominal spring constant, $k = 0.4$ N/m, resonant frequency of 70 kHz and a tip radius of 2 nm. The resulting AFM maps were used to estimate strain across different areas of the 2D membranes. From AFM traces through the membrane centre, we estimate that the membranes were globally strained by typically 0.5%. However, the strain $\varepsilon$ was distributed not uniformly but accumulated around the aperture rim[23], leading to $\varepsilon$ several times higher than away from it[23]. This yields $\varepsilon$ of a few percent around the rim. Such strain is also expected to accumulate around wrinkles in the 2D membranes, whose complex morphology cannot be attained using strain-free (bending only) deformations[24,26]. From the height ($h$) and base ($L$) of the wrinkles measured in AFM, we also estimate strain of a few percent, consistent with the above expectations.

For SEM characterisation, we used a Zeiss Gemini500 SEM, employing an In-Lens secondary electron detector, accelerating voltages of 0.5 - 2 keV and a working distance of ∼2 mm.

**Additional examples of devices studied by SECCM.** Extended Data Fig. 4 (graphene) & 6 (hBN) show further examples of SECCM and AFM maps. In all the measured devices (more than twenty 2D membranes), we observed a clear correlation between high activity areas in the SECCM maps and morphology of 2D membranes. In particular, Extended Data Fig. 6 provides an example where proton conductivity becomes sharply suppressed crossing the boundary from monolayer hBN to a multilayer region. Previously, it was shown that hBN monolayers were highly proton permeable, whereas hBN crystals of 4 or more layers in thickness exhibited indiscernible proton conductance[8]. The images of Extended Data Fig. 6 illustrate this property with nanoscale resolution across individual membranes in the same experiment. An additional interesting feature seen in the AFM maps is two wrinkles that extend along the $SiN_x$ substrate beyond individual apertures. These wrinkles display notable proton-conducting activity in the SECCM maps which occurs not only above the Nafion region but extends onto the $SiN_x$ substrate. We attribute this observation to water that fills the space between the substrate and wrinkles and thus provides a proton-conducting medium inside the wrinkles.

**Absence of defects in mechanically exfoliated 2D membranes.** Suspended membranes made from exfoliated 2D crystals have previously been characterised extensively using AFM, SEM, Raman, TEM and STM[2,5,6,8,9,25,45,46] as well as gas permeation measurements[1,2,4,5]. None of those studies could detect any structural defects in the membranes. Nevertheless, it was important to ensure that the fabrication procedures used in the present report did not lead to accidental tears, cracks or pinholes that would break the continuity of the graphene lattice and leak protons through.

The formation of wrinkles in supported thin sheets is a universal phenomenon that arises from non-uniform adhesion between the sheet and the substrate. For example, this phenomenon has been extensively studied for 2D polymers[47], and graphene is no exception. To understand the formation of wrinkles in our devices we note that graphene sheets are initially suspended over holes, rather than supported. The membranes are therefore stretched laterally because of adhesion to the holes' sidewalls and free to relax in the out-of-plane direction. In most cases, this results in wrinkle free membranes[2]. The situation changes after depositing Nafion. Adhesion to sidewalls disappears in the presence of water (as observed in ref. [48]) so that graphene is no longer stretched over the holes. The membrane therefore becomes looser, which unavoidably results in the formation of wrinkles. In addition, the now loose graphene sheets conform to the porous Nafion polymer surface, which further contributes to the wrinkling and rippling. Importantly, the wrinkles and roughness do not lead to cracks, tears or pinholes that would allow unimpeded proton permeation through them. This conclusion is supported by many experimental observations. For the sake of brevity, we describe below only three of them.

First is the Raman spectra observed for the wrinkled membranes on Nafion. Any defects in graphene leading to breakdown of its continuous crystal lattice (cracks, tears, holes or even individual vacancies) activate the so-called D peak in its Raman spectrum. The intensity of this peak increases with defect density (e.g., refs. [33,49]). Our graphene monocrystals do not exhibit any discernible D peak, which allows us to put an upper bound on the atomic-scale defect density as $\sim 10^9$ cm$^{-2}$. This translates into no more than 10 single-atom vacancies for our entire membranes of 2 μm in diameter (e.g., refs. [8,9,45]). In contrast, the reported wrinkles are hundreds of nanometres long and, if there were any breakdown of crystallinity along them, an intense D peak would also be apparent. Occasionally, we found devices with accidental cracks formed during fabrication, and those exhibited a strong D peak. They were discarded. All the devices reported in the manuscript had no discernable D peak (Extended Data Fig. 1c). Also note that the found upper bound of ~10 atomic-scale defects in our devices cannot possibly explain the observed proton conductance. Indeed, our SECCM maps typically reveal ~100 active pixels and, to provide their proton conductance, not individual vacancies but large multi-atom pinholes would be required. This would lead to a very intense D peak, easily observable experimentally.

The second piece of evidence that rules out lattice defects in our membranes comes from measurements using liquid electrolytes (refs. [6,8]). These experiments have found similar proton conductivity as in devices measured using Nafion. Unfortunately, we cannot remove Nafion after measurements but we could remove electrolytes. In the latter case, the membranes did not show any D peak or any damage under AFM or SEM, which demonstrates that the membranes were not damaged during proton conductivity measurements. Because the conductivity using electrolytes is the same as in the case of Nafion, we can safely conclude that Nafion does not damage graphene membranes either.

Finally, gas impermeability of our graphene-Nafion devices also proves the absence of defects induced by deposition of Nafion. Unlike graphene, which is completely impermeable to helium, thin Nafion films (after graphene was removed) exhibited notable helium leakage. This was measured using a He-leak detector that allowed us to resolve flows as low as $10^7$ atoms s$^{-1}$. Nafion coated graphene devices with an accidental crack display notable helium permeability, whereas undamaged devices remain leak free, despite the presence of wrinkles.

**SECCM of CVD graphene.** For these measurements, cm-scale pieces of CVD graphene (grown on Cu) were transferred onto a Nafion N212 film as reported previously[50]. To this end, the Cu foil that was covered on both sides with graphene was first exposed from one side to oxygen plasma, which removed graphene from that side. The CVD graphene remaining on the other side was then hot-pressed against the Nafion film, and the Cu foil was etched away in an ammonium persulfate solution. The resulting graphene-on-Nafion stack was left in deionised water for days to remove etchant residues. For SECCM measurements, cm-sized graphene-on-Nafion samples were fixed[17] to the Pt electrode (as described above) and characterised using the same procedures as for micrometre-sized 2D crystals.

Extended Data Fig. 5 shows that proton currents detected by SECCM for our CVD graphene were below 100 pA. There were no spots with very high currents similar to those observed for lower-quality CVD graphene devices[15,17]. Statistics of the currents collected over large areas (Extended Data Fig. 5c) can be separated into two groups. The first group of pixels displays currents of 0.1–10 pA; this is similar to those found in mechanically exfoliated graphene reported in the main text. The second group of pixels exhibits a normal distribution with the mode at ~20 pA, which is ~10 times higher than currents in the first group. AFM and scanning electron microscopy images revealed that the higher activity areas resulting in the second group came mostly from grain boundaries (Extended Data Figs. 5b-e). The higher permeability for these pixels can be attributed to multiple crystal-lattice defects present in grain boundaries (e.g., 8-atom rings that are highly proton-conductive[7,20] or even bigger defects). We have also found that grain boundaries in CVD graphene were often accompanied by local corrugations (cf. Extended Data Figs. 5d&e) with $h \approx 60$ nm and $L \approx 500$ nm, which may also have contributed to their proton permeability ($h/L \approx 0.1$).

The experiments described in this section provide important insights into the large variability in proton permeability reported in the literature for CVD-graphene films[51]. Even excluding gross defects (e.g., cracks and tears), which sometimes are prevalent in CVD graphene films[17], nanoscale pinholes can result in isolated hot spots with proton currents[17] up to 1 nA. For higher-quality CVD graphene without such defects, proton conductivity is likely to be dominated by grain boundaries. Even in the latter case, considerable variability in proton permeability is expected because of different grain sizes, depending on growth conditions, so that graphene films with smaller grains and thus higher density of grain boundaries would exhibit higher proton conductivity, in agreement with the previous report[7].

**DFT calculations.** We used the projector augmented wave method[52] implemented in Vienna ab-initio Simulation Package (VASP)[53] to model pseudopotentials of protons, and C and H atoms. The exchange-correlation potential was taken into account by considering the generalised gradient approximation (GGA) within the Perdew-Burke-Ernzerhof (PBE) form[54]. The weak van der Waals forces between graphene and proton were also included by using the DFT-D2 method of Grimme. For geometry optimisations, a kinetic energy cut-off of 500 eV was used for the plane-wave basis. The convergence criterion of the total force on each atom was reduced to $10^{-5}$ eV/Å and the convergence criterion for the energy was set at $10^{-6}$ eV. For calculating the proton barrier, we used the proton pseudopotential from the hydrogen atom and then removed an electron from the whole system.

The flat and rippled graphene were simulated as a relatively large circular-shaped crystal consisting of 150 carbon atoms (~22 Å in size), which was sufficient to prevent proton-proton interactions between neighbouring supercells. The carbon cells were isolated with a vacuum gap larger than 10 Å, which ensured the absence of edge-to-edge interactions. The ripples were modelled by fixing the out-of-plane positions of carbon atoms so that the crystal forms a Gaussian profile of height $h$ (Extended Data Fig. 7). The atoms were allowed to relax in-plane. Interatomic distances for atoms near the ripple top ($d_{cc}^{stained}$) were compared with that of the flat strain-free structure ($d_{cc}^0$) and the amount of biaxial strain was calculated as $\varepsilon = (d_{cc}^{stained} - d_{cc}^0)/d_{cc}^0$. To calculate barriers for strained flat graphene (without ripples), in-plane position of carbon atoms were obtained by applying biaxial strain. Extended Data Fig. 7b shows the energy barriers $E$ found for the three cases.

We calculated the total energy of the proton-graphene system as a function of the position of the proton in the direction perpendicular to the centre of the hexagonal ring in the graphene crystal lattice (Extended Data Fig. 7). Our calculations showed that the proton became physisorbed at ~1 Å away from the graphene lattice, which corresponded to the minimum energy of the system. The maximum was reached when the proton was in the middle of the hexagonal ring. The barrier $E$ for proton permeation is calculated by subtracting the minimum energy from the maximum one. For the case of flat unstrained graphene, the energy barrier found using these approximations is ~1.37 eV, in good agreement with the earlier theory[14]. Because various approaches used to calculate $E$ yield a rather large spread in the predicted values[10,11,13,14,18] and the exact value of the energy barrier for flat monolayer graphene remains debateable[14], here we avoid this uncertainty by focusing on relative changes in $E$ which are arising from strain and curvature. Finally, note that $E$ is expected to vary across membranes becoming lower around wrinkles and ripples and higher in flatter and unstrained areas. Because this strain is mostly random, it is reasonable to expect that in the first approximation the distribution of $E$ is normal, that is, Gaussian. Because proton currents depend exponentially on $E$, their distribution should then be log-normal, which is consistent with our SECCM observations.

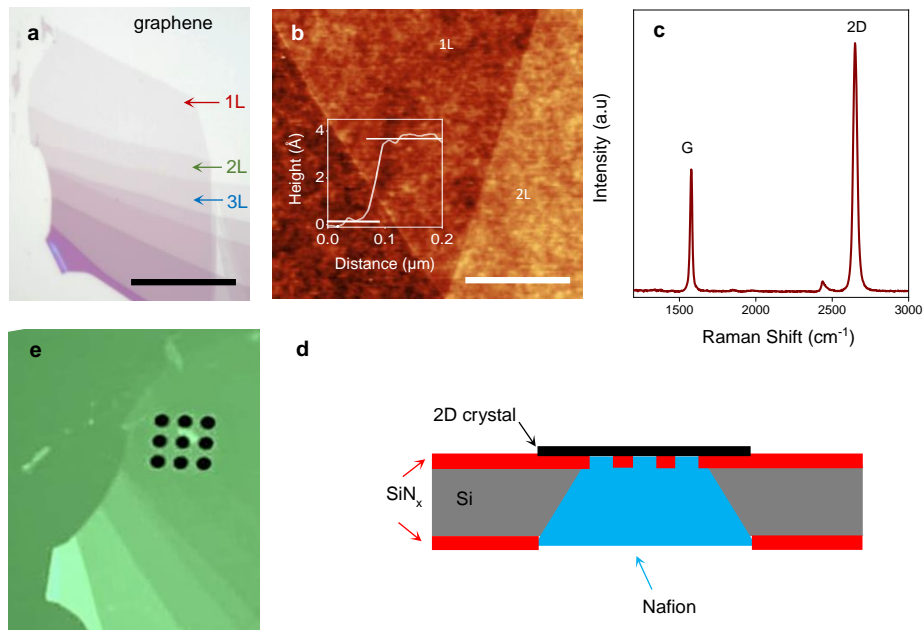

**Extended Data Fig. 1 | Experimental devices. a,** Optical image of a typical flake obtained by mechanical exfoliation. The regions that are 1-, 2-, and 3- layers thick are marked with red, green and blue arrows, respectively. Scale bar, 20 µm. **b**, AFM image of a typical graphene flake placed on an oxidized Si wafer. The inset shows the step profile corresponding to one graphene layer. Scale bar, 1 µm. **c**, Raman spectrum for our typical graphene membrane. **d**, Schematic of our experimental devices. The 2D crystal is suspended over several apertures etched into a freestanding $SiN_x$ film. The back side is coated with Nafion that is electrically connected to a Pt electrode. The top side is probed by SECCM. **e**, Optical micrograph of one our devices where the flake shown in panel **a** was transferred on its top as a membrane. The apertures (2 µm in diameter) are visible as black circles.

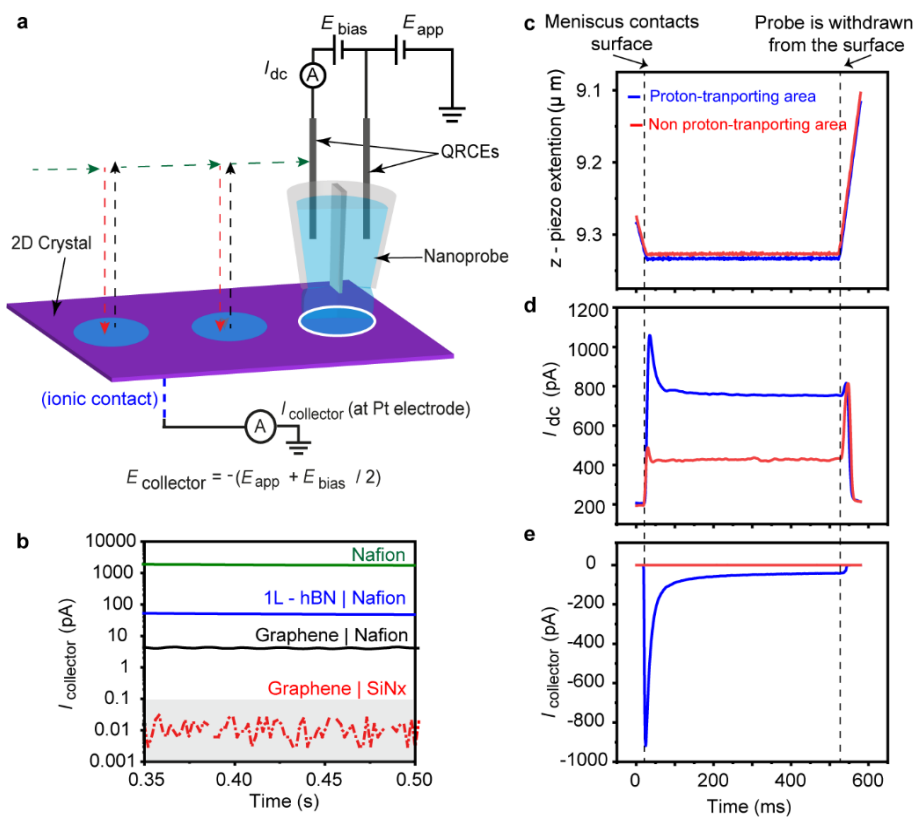

**Extended Data Fig. 2 | SECCM measurements. a**, Their basic principle. A nanopipette with two reservoirs filled with 0.1 M HCl and fitted with identical quasi-reference electrodes (QRCEs) is used to inject protons through 2D crystals. A voltage applied between the QRCEs produces an ion current $I_{dc}$ that is used as a feedback signal. Another potential ($E_{app}$) applied to one of the QRCEs determines the overall potential used to inject protons. Dashed arrows illustrate the scanning protocol: the probe approaches the crystal (red) and makes contact over the area marked by blue circle, it is then retracted (black arrows) and moved to another position (green). **b**, Steady state SECCM currents collected from various settings: green, without any crystal covering Nafion; blue, monolayer hBN on top of Nafion; black, monolayer graphene on Nafion; and red, the same graphene monolayer on $SiN_x$ away from apertures. This panel demonstrates that our experimental setup can probe proton conductivity of 2D materials within a 4-decade current range. **c**, Example of distance changes between the nanopipette and the sample during a typical scan obtained above proton-conducting (blue) and proton-blocking areas (red). **d**, $I_{dc}$ - $t$ characteristics measured simultaneously show that $I_{dc}$ accurately detects the moments when meniscus wetting and dewetting occur. **e**, Corresponding $I_{collector}$-$t$ characteristics (same colour coding). Dotted lines in panels **c-e** mark the moments when the probe wets and retracts from the surface.

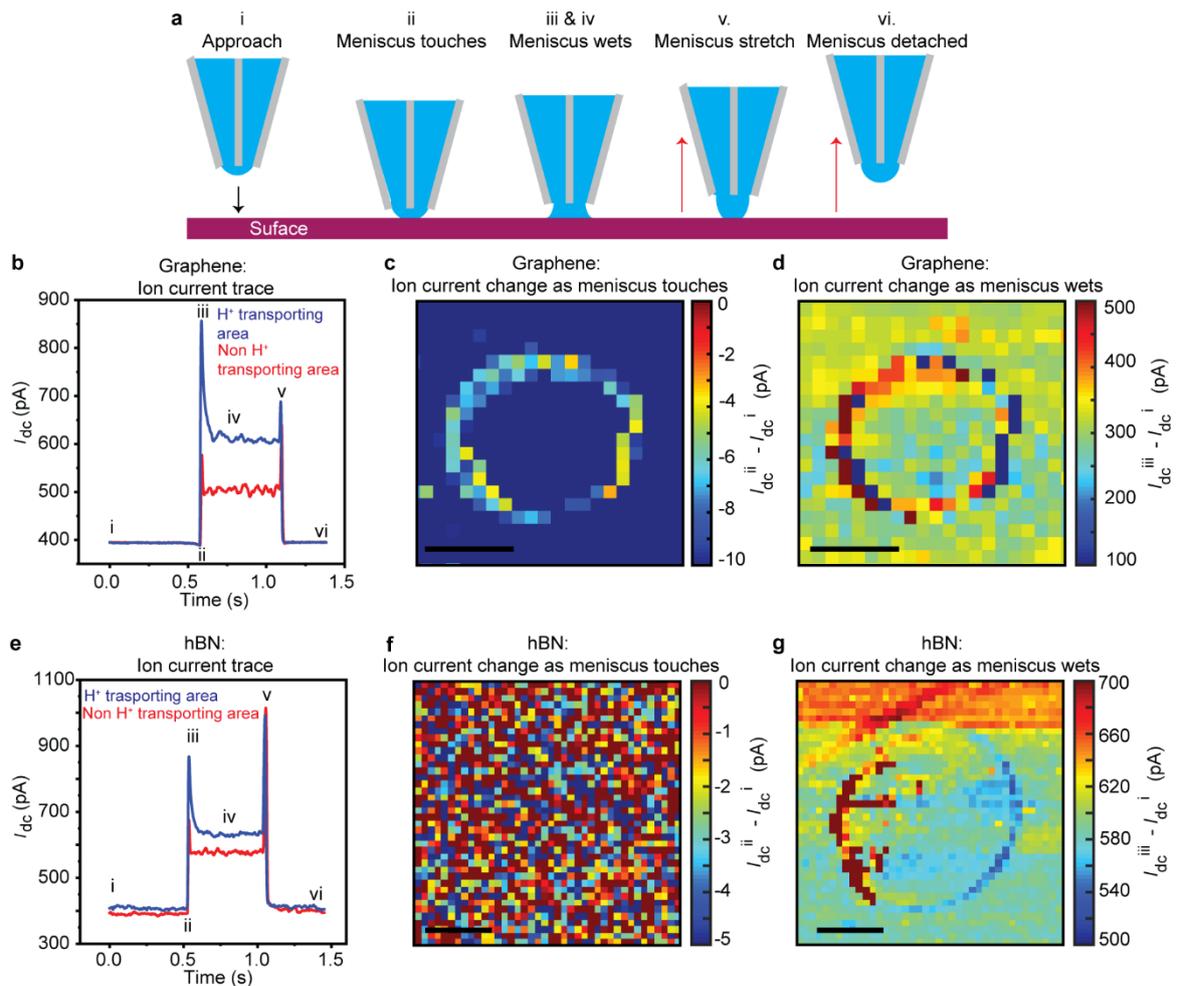

**Extended Data Fig. 3 | Meniscus-surface contact**. **a,** Schematic of the approach and retract stages for our scanning protocol. **b**, Example of $I_{dc}$ - $t$ characteristics (ion current between the two electrodes in the nanopipette) for proton conducting (blue) and non-conducting (red) areas in a graphene device. **c**, Map showing changes in $I_{dc}$ during step *ii* with respect to the current in step *i* ($\Delta I_{dc} = I_{dc}^{ii} - I_{dc}^{i}$) for the graphene device in **b**. **d**, Map of $\Delta I_{dc} = I_{dc}^{iii} - I_{dc}^{i}$ for the same device. **e-g**, Same as in panels **b**-**d** but for a monolayer hBN device. Scale bars, 1 μm.

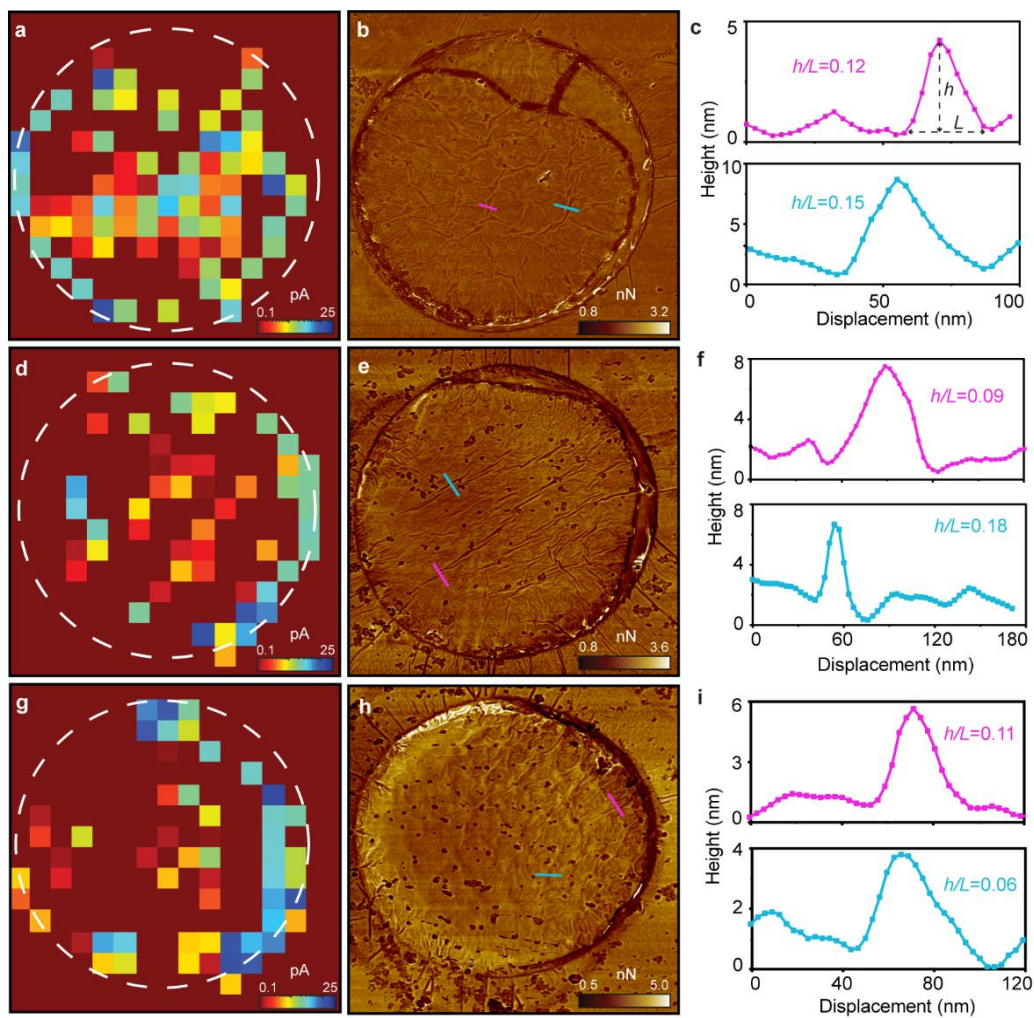

**Extended Data Fig. 4 | Additional examples of our proton-transport measurements. a,d,g,** SECCM maps for 3 graphene devices. White dashed circles mark the 2 μm apertures in $SiN_x$. **b,e,h,** AFM adhesion maps for the devices shown on the left panels. **c,f,i,** Height profiles for some wrinkles marked on the corresponding AFM maps (colour coded).

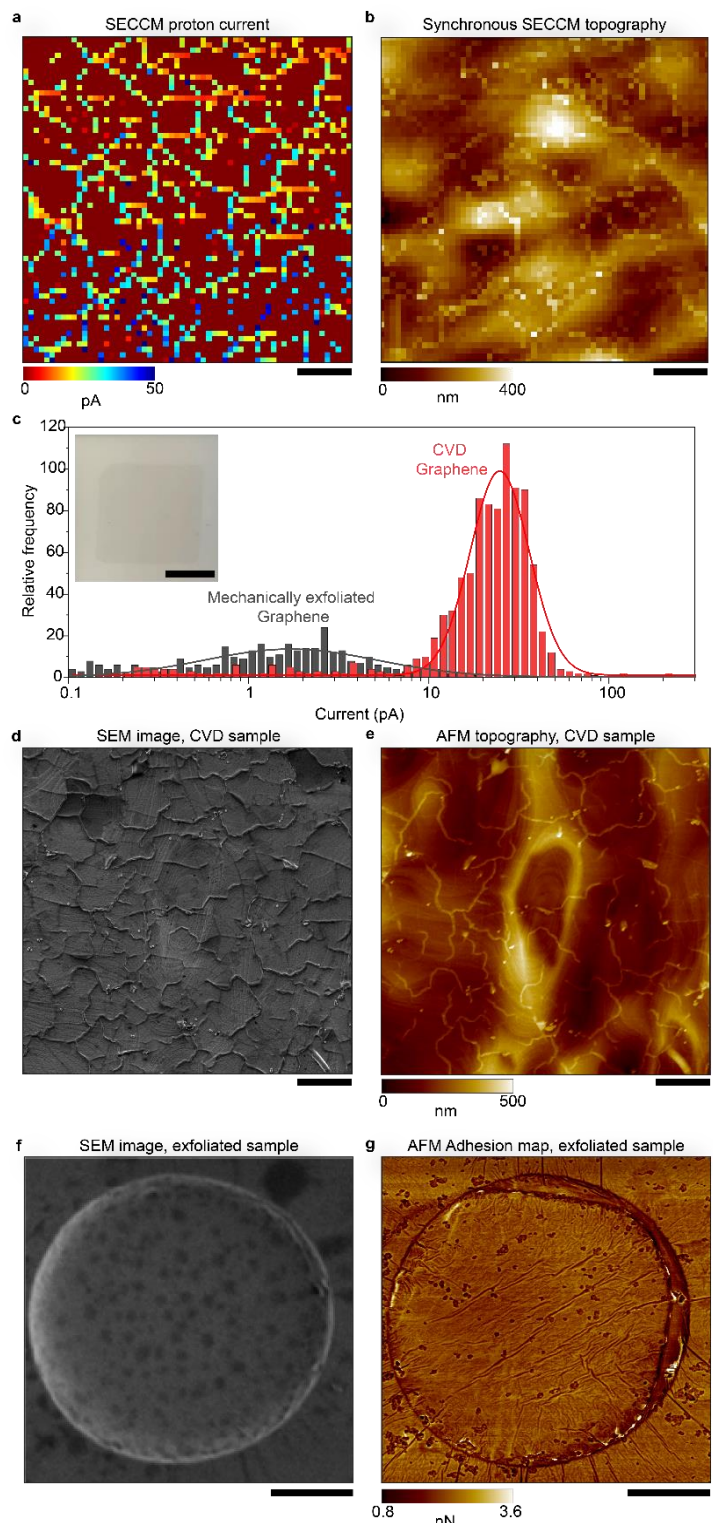

**Extended Data Fig. 5 | Proton transport through CVD graphene. a**, SECCM map for a typical region within our CVD device. The map shows proton-permeation through a graphene area of ~120×120 μm². No single pixel with current above 100 pA was found. **b**, Topography of the area collected simultaneously with the SECCM scan in panel **a**. **c**, Statistics for steady-state proton currents from panel **a** (red bars) and for mechanically exfoliated graphene (grey bars). The CVD graphene data show a log-normal distribution with the peak at ~20 pA and a long tail that matches typical proton currents observed for mechanically exfoliated graphene. Solid curves, best Gaussian fits to the data. Inset, optical image of CVD graphene on Nafion. Graphene is visible as a slightly darker square. Scale bar, 1

cm. **d**, Scanning electron micrograph of a typical area of this CVD graphene sample. Grain boundaries in the polycrystalline film are clearly visible. **e**, AFM image of the same area as shown in panel **d**. The scale bars in panels **a**, **b**, **d** and **e** are 20 µm. **f**, Scanning electron micrograph of a typical mechanically-exfoliated graphene membrane coated with Nafion from the back side. Areas with different (brighter and darker) contrast arise because of the polymer's porosity, which is evident under higher magnification, as previously reported[50]. **g**, AFM adhesion map for the device in panel **f**. Scale bars in panels **f** and **g**, 1 µm.

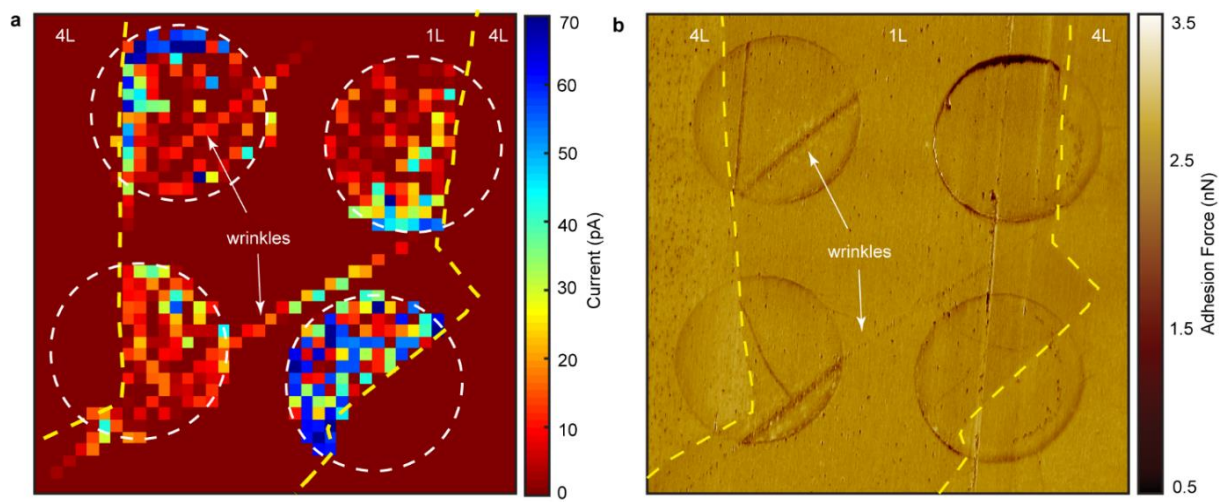

**Extended Data Fig. 6 | Additional SECCM maps for hBN. a,** Four apertures covered with mono- and tetra- layer hBN. Dashed white circles mark the apertures. Their diameter, 2 µm. The yellow dashed lines show the border between areas covered with 1L and 4L hBN. The device in the lower right corner is the one discussed in Fig. 2 of the main text. **b**, AFM force maps for the devices in **a**. Two wrinkles that facilitate proton transport above the silicon-nitride substrate (away from the apertures) can be seen to originate from the two left devices.

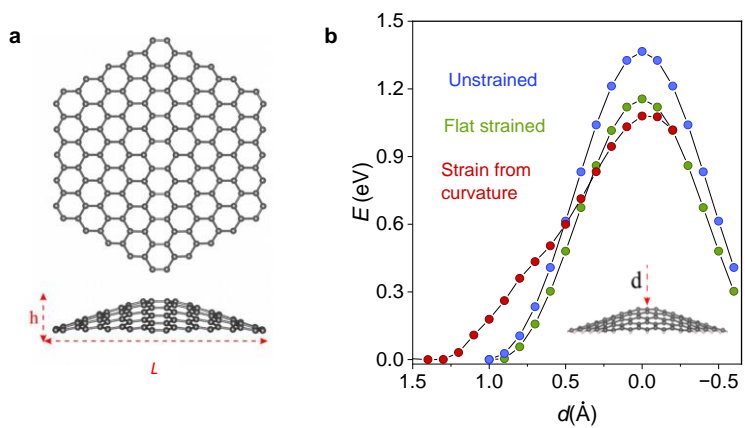

**Extended Data Fig. 7 | DFT calculations. a,** Simulated structure of rippled graphene. Its plan and side views (top and bottom, respectively). **b**, Energy barrier for proton penetration for strain-free graphene (blue), flat graphene with biaxial strain of 3.3% (green) and rippled graphene with strain arising from curvature ($h/L = 0.10$, $\varepsilon = 3.3\%$). Calculations for $E$ were done using 0.1 Å steps for the distance $d$ from graphene as shown in the inset.